\documentclass[authoryear,preprint, 3p]{elsarticle}

 \usepackage{amssymb}
\usepackage[table]{xcolor}
\usepackage{enumitem}
 \usepackage{amsthm}
 \usepackage{amsmath}
 \usepackage{natbib}
 \usepackage{tikz}
 \usepackage{graphicx}
 \usepackage{epstopdf}
 \usepackage{xurl}
 \usepackage{hyperref}
 \usepackage{multirow}
 \usepackage{kpfonts}
 \usepackage{lineno}
 \theoremstyle{plain}
 \newdefinition{definition}{Definition}

 \newdefinition{remark}{Remark}
 \newdefinition{property}{Property}
  \newdefinition{example}{Example}
\usepackage{algorithm}
\usepackage[noend]{algpseudocode}

\usepackage{caption}
\usepackage{subcaption}
\newcommand{\abs}[1]{\left|#1\right|}
\usepackage{placeins}
\definecolor{lincolngreen}{rgb}{0.11, 0.35, 0.02}
\definecolor{asparagus}{rgb}{0.53, 0.66, 0.42}

\newcommand{\ep}{\textcolor{black}}

\newcommand{\R}{\mathbb{R}}

\journal{arXiv}

\begin{document}

\begin{frontmatter}



\title{COBASE: A new copula-based shuffling method for ensemble weather forecast postprocessing}


\author[TUe]{Maurits FLOS}
\author[KNMI]{Bastien FRAN\c COIS}
\author[Geo]{Irene SCHICKER}
\author[KNMI]{Kirien WHAN}
\author[TUe]{Elisa PERRONE\footnote{Corresponding author, email \url{e.perrone@tue.nl}}}

\address[TUe]{Department of Mathematics and Computer Science, Eindhoven University of Technology, Groene Loper 5, 5612 AZ, Eindhoven, the Netherlands}

\address[KNMI]{Royal Netherlands Meteorological Institute, Utrechtseweg 297, 3731 GA De Bilt, the Netherlands}

\address[Geo]{GeoSphere Austria, Hohe Warte 38, A-1190 Vienna, Austria}

\begin{abstract}
Weather predictions are often provided as ensembles generated by repeated runs of numerical weather prediction models. These forecasts typically exhibit bias and inaccurate dependence structures due to numerical and dispersion errors, requiring statistical postprocessing for improved precision. A common correction strategy is the two-step approach: first adjusting the univariate forecasts, then reconstructing the multivariate dependence.
The second step is usually handled with nonparametric methods, which can underperform when historical data are limited. Parametric alternatives, such as the Gaussian Copula Approach (GCA), offer theoretical advantages but often produce poorly calibrated multivariate forecasts due to random sampling of the corrected univariate margins.
In this work, we introduce \textit{COBASE}, a novel copula-based postprocessing framework that preserves the flexibility of parametric modeling while mimicking the nonparametric techniques through a rank-shuffling mechanism. This design ensures calibrated margins and realistic dependence reconstruction.
We evaluate COBASE on multi-site 2-meter temperature forecasts from the ALADIN-LAEF ensemble over Austria and on joint forecasts of temperature and dew point temperature from the ECMWF system in the Netherlands. Across all regions, COBASE variants consistently outperform traditional copula-based approaches, such as GCA, and achieve performance on par with state-of-the-art nonparametric methods like SimSchaake and ECC, with only minimal differences across settings.
These results position COBASE as a competitive and robust alternative for multivariate ensemble postprocessing, offering a principled bridge between parametric and nonparametric dependence reconstruction.
\end{abstract}


\end{frontmatter}

\section{Introduction}
The forecasting skills of the physical numerical weather prediction (NWP) models have been consistently improved in recent years \citep[e.g., ][]{Bauer2015}. Efforts in research have focused on enhancing physical parameterizations, assimilating non-conventional data, and boosting spatial resolution. Despite the progress made, deterministic and probabilistic NWP forecasts still show systematic errors and underdispersion. Potential causes of errors are imperfect measurements \citep[which also serve as initial conditions, e.g., ][]{Cardinali2009}, imprecise parametrization of physical equations \citep[e.g., ][]{Bechtold2014} and inherent unpredictability due to chaotic dynamics in atmospheric flows \citep[e.g., ][]{Lorenz1963}.
To account for such model errors, calibration approaches based on statistical or machine learning \ep{postprocessing} techniques have been developed in the last years: see, for example, \cite{Gneiting2005, Raftery2005, Moeller_13, Schefzik2013, Taillardat2016, Schefzik2016, Scheuerer2018, Rasp2018, Lerch2020, Vannitsem2021}. The standard statistical methods for the \ep{postprocessing} of raw ensemble forecasts are Ensemble Model Output Statistics (EMOS) \citep{Gneiting2005} and Bayesian Model Averaging (BMA) \citep{Raftery2005}. These techniques work well in practice, but are designed to improve predictions for only one variable at a time, e.g., for a fixed weather variable at a single location and specific lead-time. As a result, such methods fail to maintain the inter-variable and spatio-temporal dependencies of the forecasts \citep[e.g., ][]{Schefzik2016}. This poses challenges for forecasting high-impact weather events such as floods \citep{Schaake2010}, heat waves \citep[e.g., ][]{Baran2020}, or the onset of hurricanes, where numerous weather factors intricately interact. To address this problem, multivariate postprocessing methods have been developed. The most popular techniques are based on \emph{empirical copulas} \citep[e.g.,][]{Schefzik2013, Lakatos2023}. Namely, these are two-step approaches that use standard EMOS or BMA to correct the univariate forecast and a reshuffling technique to reconstruct the lost dependencies in the samples that are drawn from the univariate distributions. There are several ways to sample from the post-processed distribution, including drawing random samples, uniform quantiles, or stratified samples. Uniform quantiles or stratified samples generally perform better than random samples, particularly for extremes \citep{Wilks2015, hu2016stratified}. Examples of reshuffling methods are \emph{Schaake Shuffle} \citep{Clark2004}, \emph{Ensemble Copula Coupling} (ECC) \citep{Schefzik2013}, and \emph{Sim Schaake} \citep{Schefzik2016}. The common ground of these methods is that the dependence structure can be obtained from a reference sample, which can be based on past observations (for Schaake Shuffle and Sim Schaake) or on the raw forecasts (for ECC). However, these methods have some important limitations as they require \ep{on sufficiently representative} past observations or a reliable, exchangeable ensemble system that simulates a realistic dependence structure to be effective. These conditions are not always satisfied as is the case for the ALADIN-LAEF system used in Austria \citep{Wang2011}. In such cases, \emph{multivariate parametric approaches} that do not require extensive data archives nor do not assume exchangeability, could be better suited for multivariate \ep{postprocessing}.

Multivariate parametric approaches for the statistical postprocessing of weather forecasts already exist \citep[e.g.,][]{Moeller_13, Schefzik2018, jobst2024gradient, Lerch2020}. Parametric methods with \emph{advanced copula models} improve the dependence structure of the corrected sample but often fail to maintain a suitable marginal calibration as they can only be used with random sampling which might perform poorly compared to uniform quantile sampling \citep{Wilks2015}. 
This is a problem that puts copula-based methods at a disadvantage when compared to \ep{nonparametric} two-step approaches which can be applied to any univariate corrected sample. 
We believe that this limitation has hindered the potential of multivariate parametric postprocessing in both academic research and operational use. 

This paper introduces a novel approach aimed at fully harnessing the potential of two-step parametric correction techniques. Specifically, we introduce a new copula-based shuffling technique (COBASE) that leverages the adaptability of parametric copulas while preserving the marginal calibration. We present the methodology through two case studies, similar to those examined in \cite{Perrone2020} and \citet{whan2021novel}. 
Our attention is directed towards spatial dependencies in 2-meter temperature predictions using the ALADIN-LAEF system in Austria and inter-variable and spatial dependencies in 2-meter temperature and dew-point temperature predictions the Netherlands. 
The case studies demonstrate the applicability of our method in different climates, with different numerical weather prediction systems (regional and global) and number of dimensions. Our approach significantly outperforms the state-of-the-art multivariate parametric methods, and it provides a viable way to enhance the performance of any parametric copula-based method in the \ep{postprocessing} context.

The paper is organized as follows. In Section~\ref{sec:data}, we present the two numerical weather prediction systems and our case study settings. In Section~\ref{sec:methods}, we review existing \ep{nonparametric} and parametric methods for \ep{postprocessing}, discuss the limitations of current approaches, and introduce COBASE, our new methodology. Section~\ref{sec:results} tests COBASE in our case studies. Finally, we draw conclusions and end with a general discussion in Section~\ref{sec:conclusions}.

\section{Data}
\label{sec:data}

\subsection{Case study 1: spatial dependencies in Austria}

First, we consider a case study based on the ALADIN-LAEF model \citep{Wang2011}, which is the predecessor of the currently operational Austrian ensemble forecasting system C-LAEF \citep{wastl2021}. 
The ALADIN-LAEF system has been optimized for the Austrian atmsopheric conditions and was operational from 2013 to 2022. 
The migration to the successor model C-LAEF was carried out in parallel in 2022 and put into operation in 2023. Although C-LAEF replaced ALADIN-LAEF operationally in 2023, historical raw forecasts from C-LAEF remain scarce. Therefore, ALADIN-LAEF data were used in this study.

The ALADIN-LAEF ensemble included a total of 17 ensemble members, comprising 16 perturbed members that are guided by 16 ECMWF-EPS members, along with one control run. 
ECMWF members acted as boundary conditions for ALADIN-LAEF, which means that each ALADIN-LAEF member was based on an ECMWF member. 
However, the physical parameterization within the ALADIN-LAEF ensemble varied, and the initial observations were perturbed in distinct ways. 
As a result, ALADIN-LAEF was not an exchangeable system and did not contain any of the original and unmodified members of the ECMWF. 
ALADIN-LAEF was set to run with a horizontal resolution of 10.9 km using a Lambert conformal projection. For the vertical structure, it incorporated 45 hybrid pressure-coordinate levels aligned with the terrain, averaging nine levels in the lowest 1000 km from the surface. 
The system underwent initialization twice a day at 0000 and 1200 UTC, offering hourly lead-times extending up to 72 hours into the future. The ALADIN-LAEF forecast data was subjected to bilinear interpolation to align with the locations of the TAWES (TeilAutomatische WEtterStation) network in Austria. The TAWES network, utilized for weather observations across Austria, consists of approximately 300 sites that cover almost every altitude and climatic region of the nation. 
This research examines 2-meter temperature (T2m) data collected from specific TAWES locations.
Here, we analyze the same setup and dataset used in \cite{Perrone2020}. Specifically, we examine three groups of stations over a time span extending from January 2014 to May 2018. Each group includes three sites, demonstrating a unique feature of interest as detailed below.
\begin{enumerate}
\item \textbf{LAEF - Single Valley.} The initial group comprises three stations situated in proximity within a valley. The stations chosen are {Sonnblick (11343), Kolm Saigurn (11344),} and {Rauris (11346)}.
\item \textbf{LAEF - Mountain Peaks.} The second group includes the stations located on the tops of three mountains. We focus on {Patscherkofel (11126), Pitzal (11316),} and {Sonnblick (11343)}.
\item \textbf{LAEF - Randomly Chosen.} The third group comprises three stations that are each randomly picked from separate regions within Austria. These are {Wien Hohe Warte (11035)}, {Innsbruck (11320)}, and {Graz (11290)}.
\end{enumerate}

In \emph{Single Valley}, we note that the stations are located near each other. In contrast, the stations within the \emph{Mountain Peaks} and \emph{Randomly Chosen} categories are distributed across regions spanning hundreds of kilometers. This research aims to assess spatial dependencies while assuming that prediction horizons are fixed. In particular, the study is based on a fixed lead-time of 24h.


\subsection{Case study 2: inter-variable and spatial dependencies in the Netherlands}

For this case study, we take operational forecasts of 2-meter temperature (T2m) and dew-point temperature (DPT) from the global ensemble prediction system run by the European Center for Medium-Range Weather Forecasting (ECMWF). 
We use the 51-member \ep{ECMWF ensemble, consisting of one unperturbed control forecast and 50 perturbed ensemble members,} that are initialized each day at 12 UTC during the period 2021-2024, and take the forecasts that are valid at +24 hours. \ep{The ensemble thus comprises two exchangeable groups: the statistically equivalent perturbed members and a distinct control member.} There were several model upgrades during this time, including a change in horizontal resolution from ~18km to ~9km in 2023. The use of a rolling training window, as explained in Section~\ref{sec:methods}, reduces the impact of these model changes. 

We take observations of T2m and DPT from the KNMI network of automatic weather stations in the Netherlands, similarly to \cite{whan2021novel}. We match the forecasts to the observations with cubic interpolation. For our analysis, we consider one group as detailed below.

\begin{enumerate}[resume]
\item \textbf{ECMWF ENS.} The last group comprises six stations from the total set of stations in the Netherlands chosen to be evenly distributed across the country. Namely, we selected De Kooy (235), Schiphol (240), De Bilt (260), Groningen (280), Vlissingen (310), and Maastricht (380). 
\end{enumerate}

\section{Methods}\label{sec:methods}

This section outlines the two-step approaches currently considered state-of-the-art in multivariate statistical \ep{postprocessing}.
The key idea of these methods is to separate the correction of forecasts into two parts, one that addresses the marginal forecasts, and another that pertains the dependence structure. 
In practical terms, this involves initially applying univariate \ep{postprocessing} techniques to each margin separately, meaning that we address each weather variable, lead time, and station on an individual basis. Afterwards, we restore the missing dependencies by applying the theory of \textit{copulas}.

Copulas are cumulative joint distributions with uniform margins on $[0,1]$. The theoretical foundation of copula theory is Sklar's theorem \citep{SKLAR1959}, which states that the $d$-dimensional multivariate cumulative distribution function $F$ of a random vector $(X_1, \hdots, X_d)$ with marginals $F_1,\hdots, F_d$ can be expressed as the composition of a copula $C$ and univariate marginal distributions, i.e., $F(x_1,\hdots,x_d) = C(F_1(x_1),\hdots, F_d(x_d))$. Therefore, copulas can identify the dependence structure of a phenomenon without considering its marginal distributions. In the \ep{postprocessing} context, they can be used in the second step to reintroduce a dependence structure which gets lost in the univariate \ep{postprocessing} and sampling step. In the remainder of the section, we first introduce the univariate correction method used in our study, followed by the nonparametric and parametric copula-based approaches, and our newly proposed method COBASE. A summary of the specifications of the methods used with the corresponding labels is given in Table~\ref{tab:method_labels}.

%

\subsection{Univariate \ep{postprocessing}: Ensemble Model Output Statistics (EMOS)}
One widely adopted method for univariate correction is Ensemble Model Output Statistics (EMOS) \citep{Gneiting2005}. EMOS employs ensemble forecasts as covariates to adjust the parameters of a Gaussian distribution, aiming to correct errors in the mean as well as in the forecast uncertainty.
In this paper, we follow the same approach as in \cite{Perrone2020} and define the regression model as:
\begin{align}
y &\sim \mathcal{N}(\mu, \sigma^2)\label{equ:emos}\\
\mu &= \alpha_0 + \alpha_1 \cdot m\label{equ:emos:mu}\\
\sigma^2 &= \beta_0 + \beta_1 \cdot s^2\label{equ:emos:sigma}
\end{align}
where $(\alpha_0, \beta_0)$ and $(\alpha_1,\beta_1)$ are the corresponding intercept and slope coefficients, and $m$ and \ep{$s^2$} are the mean and variance of the ensemble, respectively. 
To ensure the scale parameter $\sigma$ stays positive, a quadratic link is utilized, which compels the regression coefficients $\beta_0$ and $\beta_1$ to be nonnegative \citep{Gneiting2005}. 
The regression coefficients are calculated by minimizing the Continuous Rank Probability Score (CRPS) through the analysis of a dataset composed of past observation and ensemble forecast pairs.
\ep{In our case study, for both systems ALADIN-LAEF and ECMWF, estimation is performed by a rolling training window of ensemble forecasts and their corresponding observations of 30 days prior to the verification date}, as originally proposed by \citet{Gneiting2005}. \ep{Parameter estimation relies on numerical optimization of the Continuous Ranked Probability Score (CRPS) \citep{hersbach2000decomposition,Gneiting2005} using the Broyden–Fletcher–Goldfarb–Shanno (BFGS) algorithm \citep{Nocedal2006}. We use the empirical, data-driven formulation of the CRPS, rather than closed-form expressions, to ensure consistency with the estimation framework used in \cite{Perrone2020} and to retain flexibility across different distributional assumptions. 
To quantify the skill of the corrected forecasts, we use scoring rules such as the CRPS, which is widely applied in the univariate setting.}
For an ensemble forecast with members $x_1, \ldots, x_M \in \mathbb{R}$\ep{, predictive cumulative distribution $F_{ens}$, and observation $y$,} the CRPS is defined as follows:
\begin{equation}
\text{CRPS}(F_{ens},y)= \dfrac{1}{M} \sum\limits_{m=1}^M |x_m - y| - \dfrac{1}{2M^2} \sum\limits_{n=1}^M\sum\limits_{m=1}^M|x_n - x_m|.
\end{equation}

While EMOS provides effective univariate corrections, it does not capture dependencies across multiple variables or locations, motivating the use of multivariate \ep{postprocessing} methods. In the following section, we review \ep{nonparametric} empirical copula-based postprocessing approaches.
%

\subsection{Nonparametric multivariate \ep{postprocessing}: Empirical-copula methods}
%
%
This section reviews the most common two-step nonparametric techniques for statistical \ep{postprocessing}. 
All of these methods are based on \textit{empirical copulas}, which offer an effective way to incorporate the dependence structure into weather forecasting problems without imposing any parametric assumptions \citep{Schefzik2015}. 
An empirical copula is essentially a discrete form of a copula $C$, known as a \textit{discrete copula}, derived from a given multivariate sample \citep{Schefzik2015}. 
As such, it encodes the dependence structure of the reference sample through the corresponding ranks. 
In weather forecasting, empirical copulas can be used to generate adjusted multivariate forecasts that integrate dependencies taken from a reference rank structure. This idea was first introduced in the Schaake Shuffle method \citep{Clark2004}, which uses the rank structure obtained from historical observations. Then, this concept was refined and new methods were developed, the most popular being ECC \citep{Schefzik2013} and Sim Schaake \citep{Schefzik2016}. ECC reconstructs the dependence according to the raw ensemble forecasts. Sim Schaake combines the ideas of ECC and \ep{Schaake Shuffle}: First, it selects past days where the ensemble forecasts are similar to the verification day and then uses the selected observations to reconstruct the dependence. 

We now formally describe the steps of all these methods. Here, we use the same notation as in \cite{Perrone2020}. 
For a mathematical description of discrete copulas and their mathematical interpretation, we refer the reader to \cite{Perrone2019} and references therein. Here we only consider one lead time and assume an ensemble system of $M \in \mathbb{N}$ members, with $d=J \cdot K$, and $J, K \in \mathbb{N}$, univariate raw margins of the form $(x_1^{(j,k)}, \ldots, x_m^{(j,k)})$, where $j \in \{1, \ldots, J\}$ is a location and $k \in \{1, \ldots, K\}$ a weather variable. 
The steps shared by Schaake Shuffle, ECC, and Sim Schaake are reported in Algorithm~\ref{alg:2-step}. 
The methods differ in how they obtain the reference dependence structure $DC_{ref}$. That is, Schaake Shuffle constructs $DC_{ref}$ from randomly selected past observations. 
ECC generates $DC_{ref}$ from the raw ensemble forecasts. 
Finally, Sim Schaake builds $DC_{ref}$ from \textit{similar} past observations. 
A crucial aspect of this procedure is the meaning of \textit{similar} in choosing past days. \cite{Schefzik2016} proposes to measure similarity by a criterion, which assigns a score to each available date. The score is determined by evaluating the raw forecasts of the verification date against the raw forecasts of historical dates. Here, we follow the approach proposed by~\cite{Schefzik2016} and select a pool of similar past days through the following criterion $\Delta$:
\begin{equation}
\label{eq:simcr}
\Delta(\mathbf{x}^t,\mathbf{x}^{t_p}):=\sqrt{\dfrac{1}{d}\sum_{l=1}^{d} \left(\bar{x}^{l,t} - \bar{x}^{l,t_p}\right)^2 + \dfrac{1}{d}\sum_{l=1}^{d} \left(s^{l,t} - s^{l,t_p}\right)^2}
\end{equation}
where $\bar{x}^{l,\tau}$ and $s_{l,\tau}$ are the empirical mean and standard deviation of the ensemble forecast $\mathbf{x}^{\tau}$ for date $\tau$, i.e., they are defined as follows:
\[
\bar{x}^{l,\tau} := \dfrac{1}{M} \sum_{m=1}^M x_m^{l,\tau}, \quad
s_{l,\tau}:= \sqrt{\dfrac{1}{M-1}\sum_{m=1}^M \left(x_m^{l,\tau} - \bar{x}^{l,\tau} \right)^2}.
\]
In addition to the construction of the reference dependence structure $DC_{ref}$, another point that can generally vary is the sampling technique used in \textbf{\ep{Step3}} of Algorithm~\ref{alg:2-step}. In \cite{Schefzik2013}, various sampling techniques have been discussed. One could simply take equidistant quantiles of step $N+1$ from the univariate corrected distribution $F_l$:
\begin{equation}
\label{Quantiles}
\tilde{x}^l_1 = F^{-1}_l \left(\dfrac{1}{N+1}\right),\; \ldots, \; \tilde{x}^l_N = F^{-1}_l \left(\dfrac{N}{N+1}\right). 
\end{equation} 
There is a general consensus in the literature that the most natural way to construct a discrete sample of size $N$ by uniform quantiles is also the most effective in this context \citep[e.g.][]{Wilks2015}. 
This is true especially when uniform quantiles are compared with randomly drawn samples. 
Due to this and in line with \cite{Perrone2020}, we use uniform quantiles in our study.

\begin{algorithm}[t!]
\caption{Two-step rank-based postprocessing methods}\label{alg:2-step}
\begin{algorithmic}[0]
\State {\bf \ep{Step1}:}  Derive a reference dependence structure $DC_{ref}$;
\State {\bf \ep{Step2}:} Apply EMOS and obtain a corrected distribution $F_l$ for each margin $l=1, \ldots, d$;
\State {\bf \ep{Step3}:} Draw a new sample of size $N \in \mathbb{N}$ from each corrected distribution $F_l,$ where $l=1, \ldots, d$;
\State {\bf \ep{Step4}:} Shuffle the corrected univariate samples according to the ranks of $DC_{ref}$.
\end{algorithmic}
\end{algorithm}

In our case studies, we ensure a fair comparison with ECC by setting the size of the ensemble at $N=17$ for the ALADIN-LAEF groups and $N=51$ for the ECMWF data.
For the Sim Schaake and Schaake Shuffle methods \ep{(in both datasets)}, we use all available observations and raw forecasts, including past and future dates \ep{after the validation date} (approximately four years), to ensure a consistent number of reference dates throughout the test set. 
\ep{Although future observations are of course not available in an operational setting, we consider it reasonable to assume that four years of past observations and forecasts can be accessed, which is why we include future dates in this analysis.}
Specifically for the Schaake Shuffle, we select observations randomly from a 14-day window before and after the forecast date to preserve the seasonal climatology. \ep{We additionally examined the performance using larger window sizes and observed no notable differences in the outcomes.}

The main properties of the empirical copula-based approaches described here are summarized in Table~\ref{tab:comparison_methods_flipped}. As emphasized in the table, the main limitations of these approaches are as follows.
\ep{
\begin{itemize}[noitemsep]
\item[(a)] The direct reliance on past observations as dependence templates for reconstructing multivariate dependence, which affects both the Schaake Shuffle and Sim Schaake. This restricts the reconstructed dependence to patterns directly observed in the historical record and implicitly assumes that suitable historical analogues exist for the current forecast situation.
\item[(b)] The additional requirement of historical raw forecasts and a sufficiently large pool of past observations/forecasts to define a meaningful similarity criterion, which specifically impacts Sim Schaake.
\item[(c)] The requirement that the sample size of the corrected forecasts ($N$) must equal the number of ensemble members ($M$), which imposes a significant limitation on the applicability of ECC.
\end{itemize}
}
Although these limitations might seem less relevant in an operational context, they may have a significant impact in practice. An example is Case Study~1 in Austria, where the \ep{postprocessing} of the operational system C-LAEF could not be performed due to the lack of historical raw forecasts and historical observations. 
Furthermore, for ALADIN-LAEF, the limited ensemble size of only $17$ members posed a substantial constraint on the application of the ECC method. Consequently, this case study demonstrates a scenario in which the shortcomings of these methods might justify a transition to alternative approaches, such as parametric multivariate postprocessing approaches. We introduce these approaches in the following section after providing an overview of how multivariate forecasts are evaluated in this study.

While the CRPS is suitable for evaluating univariate forecasts, it cannot capture dependencies across multiple variables or locations. For multivariate forecasts, the Energy Score (ES) \citep{gneiting2007strictly} is commonly used as a generalization of CRPS. For an ensemble with members $x_1, \ldots,x_M \in \mathbb{R}^d$ and observation $y \in \mathbb{R}^d$, the ES is defined as
\begin{equation}
\text{ES} = \dfrac{1}{M} \sum\limits_{m=1}^M \| x_m - y \| - \dfrac{1}{2 M^2} \sum\limits_{\nu=1}^M \sum\limits_{m=1}^M \| x_{\nu}- x_m \| \ep{.}
\end{equation}
In addition, the Variogram Score (VS) \citep{Scheuerer2015} explicitly addresses miscalibrations in the correlation structure of multivariate forecasts. In the ensemble setting, it is defined as
\begin{equation}
    VS = \sum_{i=1}^d\sum_{j=1}^d {w}_{i,j}\left(\abs{{y}_i - {y}_j}^p - \frac{1}{\ep{M}}\sum_{k=1}^{\ep{M}}\abs{x^{(k)}_i - x^{(k)}_j}^p\right)^2\ep{,}
\end{equation}
where $w_{i,j}$ are weights and $p>0$ is the order of the score. Because the VS compares pairwise differences between variables in the observation and forecast, it is particularly sensitive to errors in the dependence structure.
In our case study, we use both the Energy Score and the Variogram Score to assess and compare the performance of the different multivariate \ep{postprocessing} methods. These metrics provide complementary information, evaluating both the overall accuracy of the multivariate forecasts and their ability to represent dependence structures.
We used the variant with no weights and $p=1$ for the Variogram Scores, but we obtained consistent results for other variants as well.

\begin{table}[h!]
    \centering
    \scriptsize
    \setlength{\arraycolsep}{2pt} 
    \renewcommand{\arraystretch}{1.3} 
    \begin{tabular}{l|c|c|c|c|c|c}
    \textbf{Method} & \textbf{Approach} & \textbf{Dependence} & \textbf{\textbf{\shortstack{\ep{No direct use of} \\ \ep{past dependence} \\ \ep{templates}}}} & \textbf{\shortstack{\ep{No use of} \\ \ep{historical} \\ \ep{raw forecasts}}} & \textbf{\shortstack{Flexible \\marginal\\sampling}} & \textbf{\shortstack{No restriction\\on the \\sample size}} \\
    \hline
    \textbf{\shortstack{Sim Schaake}} & Shuffling & \shortstack{Empirical copula\\(from past obs)} & {\large \textcolor{red}{\texttimes}} & {\large \textcolor{red}{\texttimes}} & {\large \textcolor{asparagus}{\checkmark}} & {\large \textcolor{asparagus}{\checkmark}} \\
    \hline
    \textbf{\shortstack{\ep{Schaake}  \ep{Shuffle}}} & Shuffling & \shortstack{Empirical copula\\(from past obs)} & {\large \textcolor{red}{\texttimes}} & {\large \textcolor{asparagus}{\checkmark}} & {\large \textcolor{asparagus}{\checkmark}} & {\large \textcolor{asparagus}{\checkmark}} \\
    \hline
    \textbf{\shortstack{ECC}} & Shuffling & \shortstack{Empirical copula\\(from raw forecasts)} & {\large \textcolor{asparagus}{\checkmark}} & {\large \textcolor{asparagus}{\checkmark}} & {\large \textcolor{asparagus}{\checkmark}} & {\large \textcolor{red}{\texttimes}} \\
    \hline
    \textbf{\shortstack{Copula methods\\(parametric)}} & \shortstack{Parametric fit\\(copula model)} & \shortstack{Parametric\\copula model\\(rolling window)} & {\large \textcolor{asparagus}{\checkmark}} & {\large \textcolor{asparagus}{\checkmark}} & {\large \textcolor{red}{\texttimes}} & {\large \textcolor{asparagus}{\checkmark}} \\
    \hline
    \textbf{\shortstack{COBASE}} & \shortstack{Parametric fit\\(copula model)\\+ shuffling} & \shortstack{Parametric\\copula model\\(rolling window)} & {\large \textcolor{asparagus}{\checkmark}} & {\large \textcolor{asparagus}{\checkmark}} & {\large \textcolor{asparagus}{\checkmark}} & {\large \textcolor{asparagus}{\checkmark}} \\
    \hline
    \end{tabular}
    \caption{\ep{Summary of the multivariate postprocessing methods analyzed in this work and their main characteristics. Check marks indicate that a method satisfies the corresponding property, while crosses indicate that it does not.}}
    \label{tab:comparison_methods_flipped}
\end{table}

\subsection{Existing parametric multivariate \ep{postprocessing}}

The reordering strategies discussed so far are, in fact, nonparametric. Specifically, within the context of Sklar’s theorem, the copula $C$ is represented by the empirical copula $DC_{ref}$ derived from a reference dependence structure that varies for each method. Another way to proceed would be to follow a two-step parametric approach. Namely, we fit a parametric copula $C$ to the training data and use it to generate a corrected multivariate sample with prescribed marginal distributions obtained through EMOS.

The standard two-step parametric approach used in the statistical \ep{postprocessing} context is the Gaussian Copula Approach (GCA). This method was originally proposed by \cite{Pinson2012} and \cite{Moeller_13} and fits a multivariate normal distribution to the data. Specifically, past observations are used to estimate the covariance matrix of a multivariate normal distribution as follows. 
First, observations are transformed into latent standard normal values; that is, the probability integral transform is used with the univariate distributions to obtain standard uniformly distributed values. 
In mathematical terms, we transform observation \ep{${y}\in\R^d$}, where \ep{${y}_i$} has distribution $F_i$, to
\ep{\begin{equation}
    \widehat{{y}}_i=\Phi^{-1}\left(F({y}_i)\right),
\end{equation}}
where $\Phi^{-1}$ is the inverse standard normal distribution function. The values are now in an appropriate format to fit a multivariate normal distribution, where the mean is the zero vector (due to the transformation) and the covariance matrix \ep{${\Sigma}$} can be fitted. The postprocessed forecast can then be found by applying the previous transformation in reverse. 
Specifically, we take the quantiles of the probability integral transform (with standard normal distribution) of the random samples. 
In equations, we transform \ep{${Z}\sim \mathcal{N}({0},{\Sigma})$} into a post-processed ensemble forecast $X^*_1, \ldots, X_M^*$, with $X_i^*:=(X_i^{*1}, \ldots, X_i^{*d})$ for $i=1, \ldots, M$ and $l=1, \ldots, d$ obtained as follows
\ep{
\begin{equation}
    X_i^{*l} = F_i^{-1}\left(\Phi({Z}_i)\right).
\end{equation}
}
Hence, the observations are transformed into normally distributed variables that are used to fit the covariance matrix. 
This covariance matrix can be thought of as the copula parameter that models all dependencies and is used to create the forecasts. 
Therefore, GCA can be considered as a genuine copula-based approach where $C$ is the Gaussian copula. 

The copula-based parametric correction steps for \ep{postprocessing} are summarized in Algorithm~\ref{alg:2-step-cop}, which illustrates that any copula model can be used in place of the Gaussian copula. 
\ep{In both case studies, copula parameters for the parametric copula models are estimated using a rolling 30-day training window. In initial experiments, longer training windows (e.g., 50 days) were also considered but did not lead to clear performance changes. Using longer windows increases the risk of mixing different seasonal statistics in the training data, while overly short windows may adversely affect the quality of dependence estimation. We therefore adopt a rolling 30-day window to ensure consistency with the EMOS-based univariate postprocessing, so that both univariate and multivariate corrections rely on the same pool of past observations. In our study, copula parameters are estimated on each rolling training window using pseudo-observations obtained from the marginal CDF values. For the Gaussian and Archimedean copulas considered here, the copula parameters are fitted using inversion of Kendall’s tau \citep{Hofert2018}, which provides a fast and robust moment-based estimator.\\
Although parametric copula-based approaches also rely on past observations for parameter estimation, their dependence on historical data is less pronounced than for empirical shuffling methods such as the Schaake Shuffle or Sim Schaake, which directly use past observations as dependence templates. In parametric copula models, the estimated dependence structure can deviate from any individual historical realization, allowing for greater flexibility in representing multivariate dependence.
}

We notice that there has been limited research on the application of alternative copulas in the \ep{postprocessing} context, with notable exceptions such as \cite{jobst2024gradient} and \cite{Flos2022}. 
As highlighted in Table~\ref{tab:comparison_methods_flipped}, while parametric copula-based methods exhibit considerable flexibility, a key limitation lies in their reliance on random sampling.
Unlike structured techniques, such as the use of uniform quantiles, random draws provide a sparse and noisy representation of the univariate distributions. Since copula methods always generate corrected samples through random draws from a multivariate distribution, the resulting margins inherit this inefficiency. 
Consequently, univariate skill often suffers even when multivariate dependencies are well captured. In the next section, we provide evidence that this issue is substantial and would affect the performance of any multivariate parametric \ep{postprocessing} method.

\begin{algorithm}[b!]
\caption{Two-step copula-based postprocessing methods}\label{alg:2-step-cop}
\begin{algorithmic}[0]
\State {\bf \ep{Step1}:} Estimate the parameters $\alpha$ of a copula model $C_{\alpha}$;
\State {\bf \ep{Step2}:} Apply EMOS and obtain a corrected distribution $F_l$ for each margin $l=1, \ldots, d$;
\State {\bf \ep{Step3}:} Draw a sample of size $N \in \mathbb{N}$ from the multivariate corrected distribution $\mathbf{F}=C_{\alpha}(F_1, \ldots, F_{d})$.
\end{algorithmic}
\end{algorithm}

\begin{figure}[!h]
    \centering
    \includegraphics[width=0.8\textwidth]{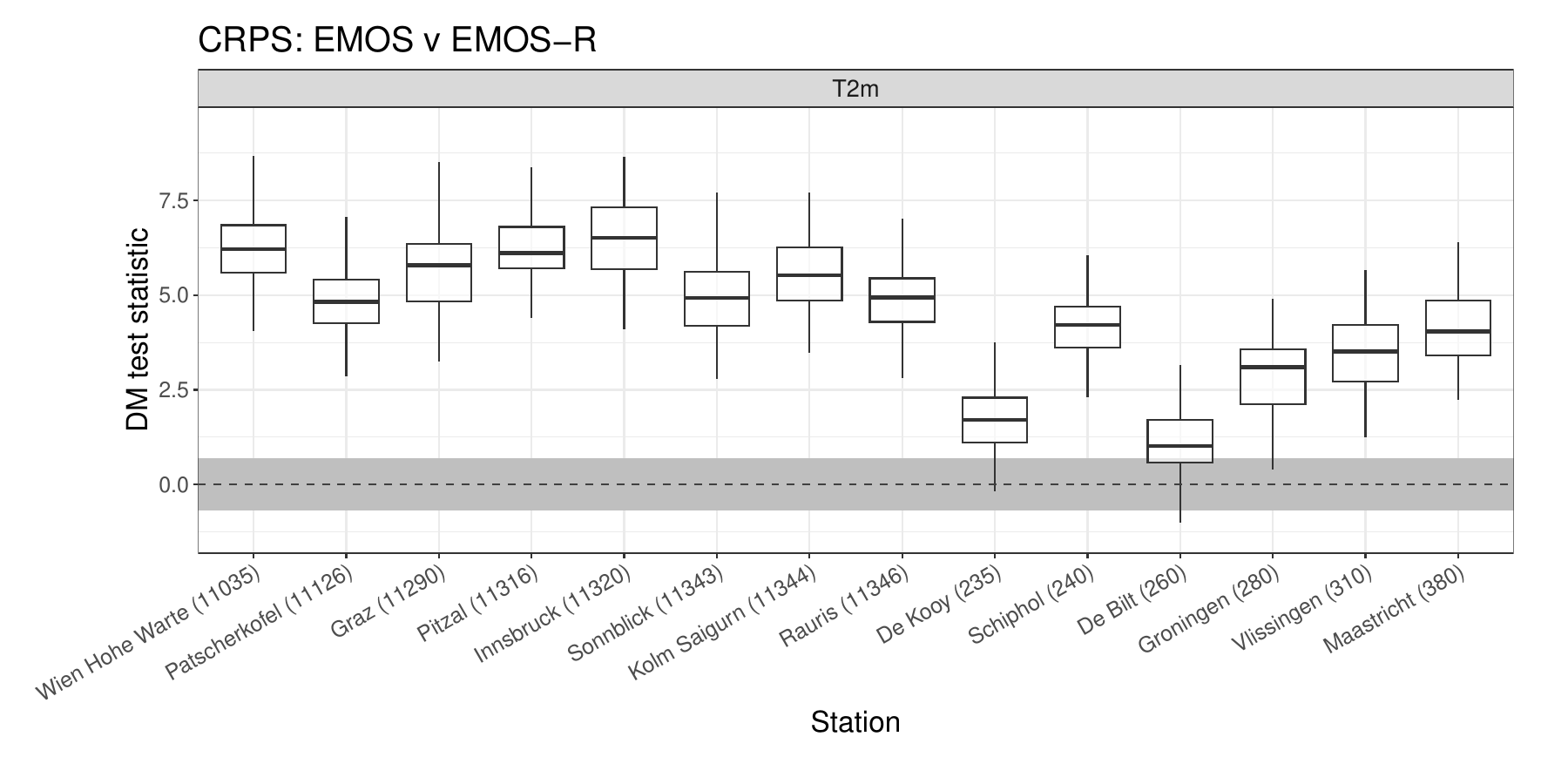}
     \caption{Boxplots of Diebold-Mariano test statistics \ep{\cite{DM1995}} for CRPS of T2m, comparing EMOS-Q against EMOS-R (baseline). Results are presented for the Austrian and Dutch stations used in this study.}
    \label{fig:crps_unshuffled}
\end{figure}

\begin{figure}[!h]
    \centering
    \includegraphics[width=0.9\linewidth]{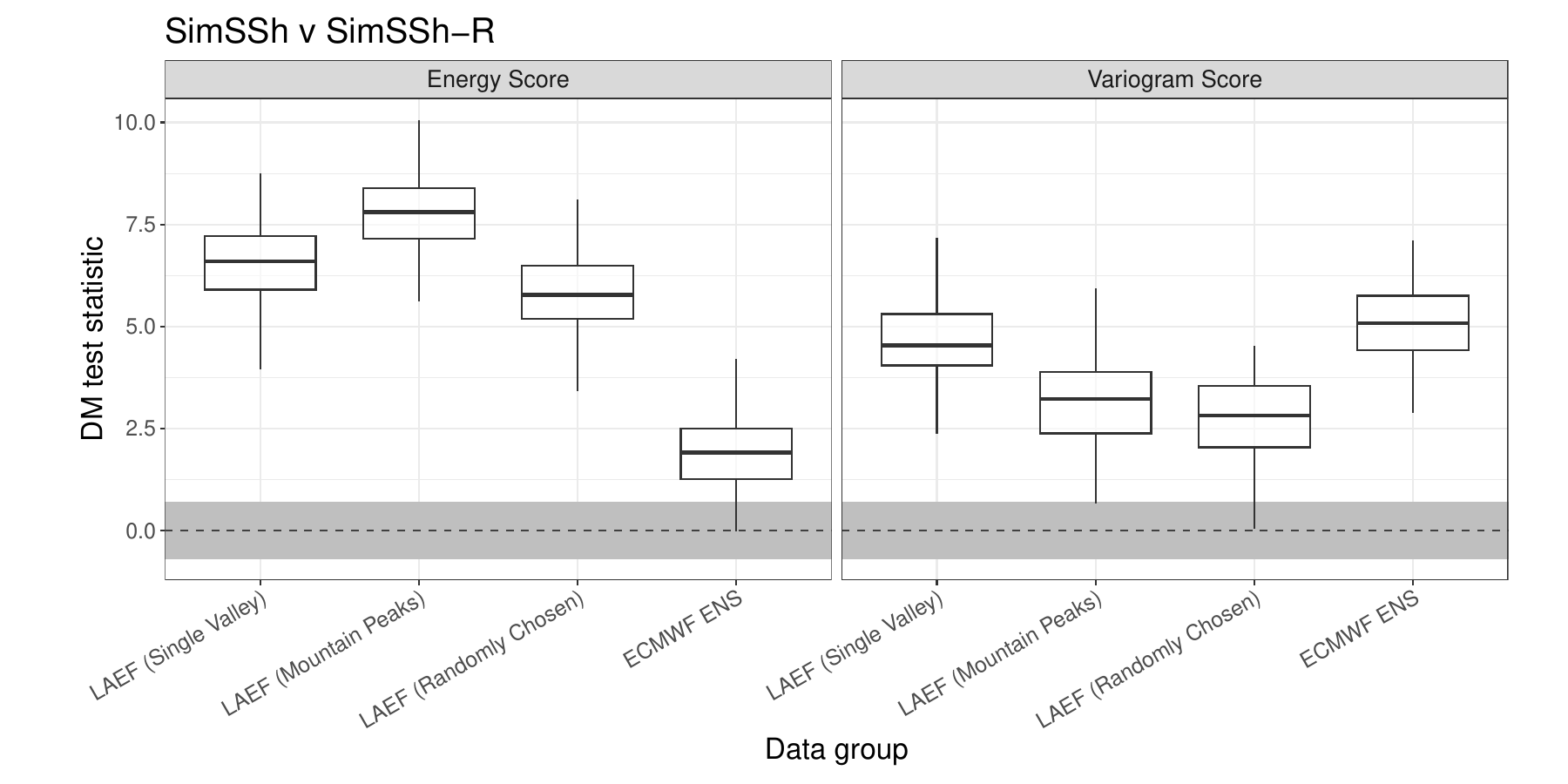}
    \caption{Boxplots of Diebold-Mariano test statistics to compare ES and VS of Sim Schaake with uniform quantile sampling (SimSSh) versus Sim Schaake with randomly drawn margins (SimSSh-R, baseline). Results are presented for T2m for the three groups of Austrian stations and for T2m and DPT for the group of Dutch stations.}
    \label{fig:multi_unshuffled}
\end{figure}

\subsubsection{Limitations of univariate random sampling in statistical \ep{postprocessing}} 

As mentioned earlier, the reliance of some univariate and multivariate \ep{postprocessing} techniques on random sampling can lead to poorly calibrated margins. To illustrate this, we compare the CRPS of T2m obtained for all stations in the study (Figure~\ref{fig:crps_unshuffled}) using EMOS-R, i.e., random draws from the corrected Gaussian distribution, as the baseline. 
Figure~\ref{fig:crps_unshuffled} shows that drawing uniform quantiles instead (EMOS-Q) consistently outperforms EMOS-R across all stations.

\ep{We further examine how univariate random sampling may affect multivariate corrections. To illustrate this effect, we focus on the Sim Schaake method, which has been shown to perform best for the ALADIN-LAEF system \citep{Perrone2020} and remains competitive for the ECMWF dataset considered here \citep{Schefzik2016}. We compare the Energy Score and the Variogram Score for Sim Schaake using different marginal sampling strategies, namely EMOS-R and EMOS-Q.} Essentially, the two methods compared in Figure~\ref{fig:multi_unshuffled} share the same dependence structure and only differ in the marginal sampling technique. 
The results show that Sim Schaake with randomly drawn margins (SimSSh-R) consistently performs worse than Sim Schaake with uniform quantile margins (SimSSh)  in both scores across all groups, demonstrating that poor univariate sampling propagates into multivariate performance.

This investigation highlights how random sampling of marginal distributions can negatively affect multivariate corrections. Since copula-based techniques fundamentally rely on random draws from parametric multivariate distributions, this limitation may \ep{negatively affect} their performance.
To overcome this limitation, we develop a new method, COBASE, which is presented in the following section.

\subsection{A new copula-based shuffling method for ensemble postprocessing: COBASE}
\label{sec:new_methods}
The proposed Copula-Based Shuffling of Ensemble (COBASE) method extends existing copula-based postprocessing techniques by introducing a shuffling of corrected univariate forecasts that replaces random marginal sampling. This approach preserves both realistic marginal behavior and the dependence structure defined by the chosen copula. The procedure is summarized in Algorithm~\ref{alg:cobase}. First, a parametric copula $C_{\alpha}$ is fitted to the data to capture the underlying dependence structure (Step S1). A random sample is then drawn from this copula to serve as the reference dependence structure $C_{ref}$ (Step S2).
Next, standard univariate postprocessing, such as EMOS, is applied to each margin $l=1, \ldots, d$ to obtain corrected distributions $F_l$ (\ep{Step3}). Independent samples of size $N$ are drawn from these corrected margins using uniform quantiles (Step S4). 
This ensures that the marginal distributions are well-calibrated, including at the extremes, avoiding the shortcomings of random draws from a parametric multivariate distribution. 
Importantly, this step is flexible: any univariate sampling technique can be used, not necessarily uniform quantiles, for example, stratified sampling, allowing additional control over the sampled distributions.
Finally, the univariate samples are shuffled according to the rank structure of $C_{ref}$ (Step S5). By doing so, COBASE preserves the dependence structure depicted by the parametric copula while leveraging the improved marginal calibration of advanced univariate sampling through, for example, uniform quantiles. 
As previously discussed, Table~\ref{tab:comparison_methods_flipped} provides an overview of the main multivariate \ep{postprocessing} methods and their key properties. 
In the table, COBASE stands out by combining the strengths of parametric and \ep{nonparametric} two-step approaches: The parametric copula captures dependence, while uniform quantile sampling ensures well-calibrated margins.
In the next section, we evaluate COBASE in two real-world case studies, comparing it with state-of-the-art methods to highlight its practical advantages.
In the two case studies, we estimate the copula parameters for COBASE using a 30-day rolling window, consistent with the EMOS procedure and the standard copula models approaches.

\begin{algorithm}[h!]
\caption{Our proposed approach - COBASE}\label{alg:cobase}
\begin{algorithmic}[0]
\State {\bf \ep{Step1}:} Estimate the parameters $\alpha$ of a copula model $C_{\alpha}$;
\State {\bf \ep{Step2}:} Draw a random sample of size $N \in \mathbb{N}$ from the copula $C_{\alpha}$ and mark it as the reference dependence structure $C_{ref}$.
\State {\bf \ep{Step3}:} Apply EMOS and obtain a corrected distribution $F_l$ for each margin $l=1, \ldots, d$;
\State {\bf \ep{Step4}:} Draw a new sample of size $N \in \mathbb{N}$ from each corrected distribution $F_l$, where $l=1, \ldots, d$;
\State {\bf \ep{Step5}:} Shuffle the corrected univariate samples according to the ranks of $C_{ref}$.
\end{algorithmic}
\end{algorithm}

\begin{figure}[!t]
    \centering
    \includegraphics[width=\textwidth]{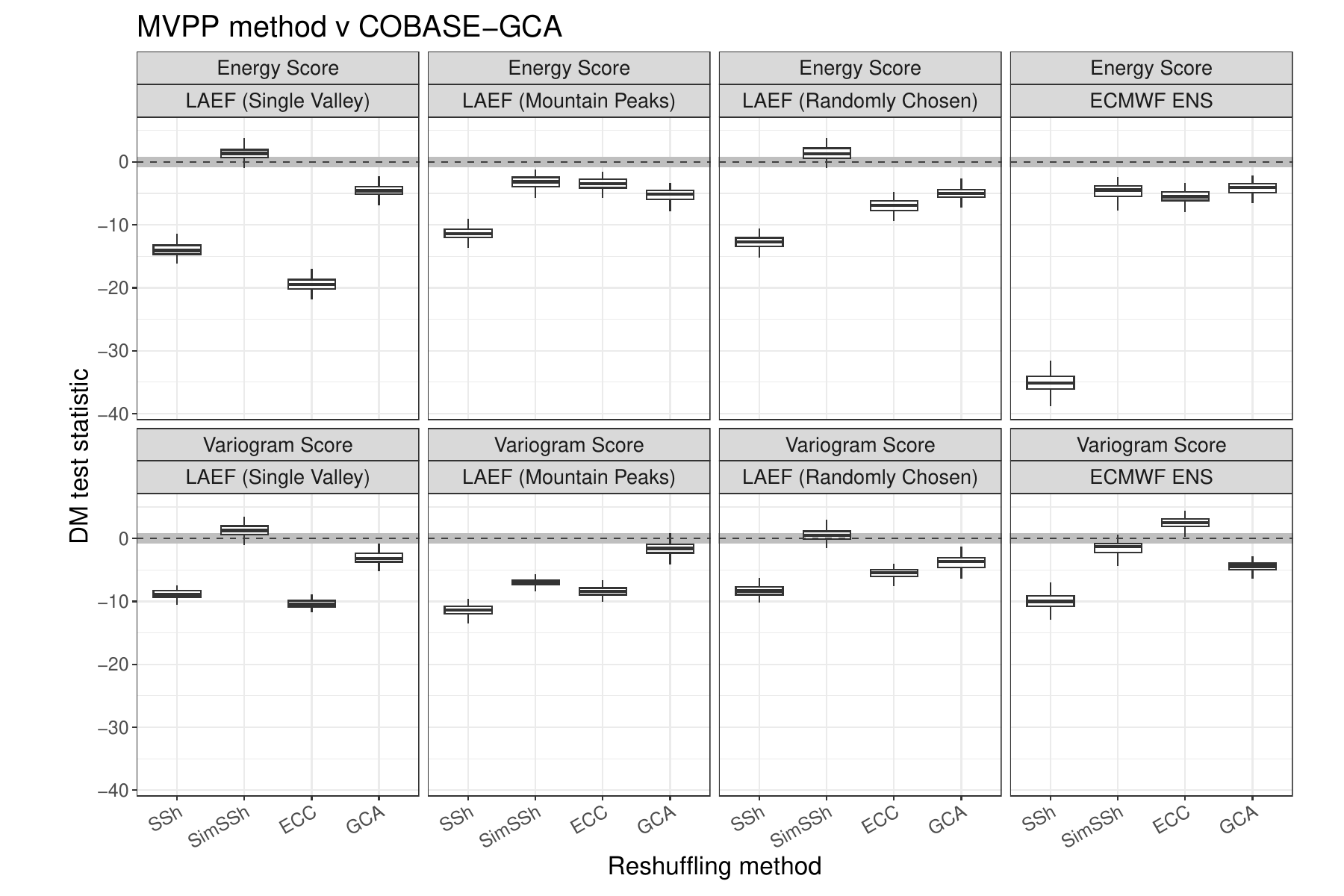}
    \caption{Boxplots of Diebold-Mariano test statistics for the case studies. Here, we compare ES and VS of Schaake Shuffle (SSh), Sim Schaake (SimSSh), ECC, and GCA with COBASE-GCA as the baseline. Values above the gray band indicate better performances compared to the baseline. Results are presented for T2m for the three groups of Austrian stations and for T2m and DPT for the group of Dutch stations.}
    \label{fig:empirical_cobase}
\end{figure}

\section{Results}\label{sec:results}
We evaluate the proposed COBASE method in comparison with state-of-the-art \ep{postprocessing} approaches on the two real-world case studies described in Section~\ref{sec:data}.
Although the focus of this work is not on univariate correction, CRPS results for 2-meter temperature across all stations are included in Table~\ref{tab:crps} for completeness. 
\ep{These results show that EMOS-Q, which uses uniform quantile sampling, outperforms methods based on random sampling on average, indicating that quantile-based approaches are generally preferable in this setting. The reported differences are based on mean CRPS values computed over the full evaluation period. Although averaging may mask the influence of individual validation days, the overall results consistently favor quantile-based sampling. Small differences in mean CRPS among methods relying on random sampling of the marginal distributions may partly reflect the variability induced by the random draws, which is a known limitation of EMOS-R and traditional copula-based approaches. This source of variability is avoided by the deterministic sampling method used in EMOS-Q.}
The main focus of this section is on multivariate forecast performance, assessed using the Energy Score (ES) and the Variogram Score (VS). 
We begin by evaluating COBASE-GCA, the variant of COBASE using a Gaussian copula. 
COBASE-GCA is compared with Sim Schaake, Schaake Shuffle, ECC, and standard GCA. 

The results, presented as boxplots of Diebold–Mariano (DM) test statistics with COBASE-GCA as reference, are shown in Figure~\ref{fig:empirical_cobase}, with the corresponding ES and VS values reported in Table~\ref{tab:ES_VS}.
These findings show that COBASE-GCA consistently outperforms standard GCA for both ES and VS for all groups of stations. 
Additionally, COBASE-GCA generally exhibits comparable performance across the different groups of stations compared to the other empirical copula-based methods. In particular, COBASE-GCA outperforms all the other methods in LAEF (Mountain Peaks). COBASE-GCA outperforms ECC and \ep{Schaake Shuffle} in LAEF (Single Valley) and LAEF (Randomly chosen), with Sim Schaake being preferable in those instances. Finally, for ECMWF ENS data, COBASE-GCA achieves the best ES, while ECC performs best in terms of VS.

\begin{figure}[!htb]
    \centering
    \includegraphics[width=\textwidth]{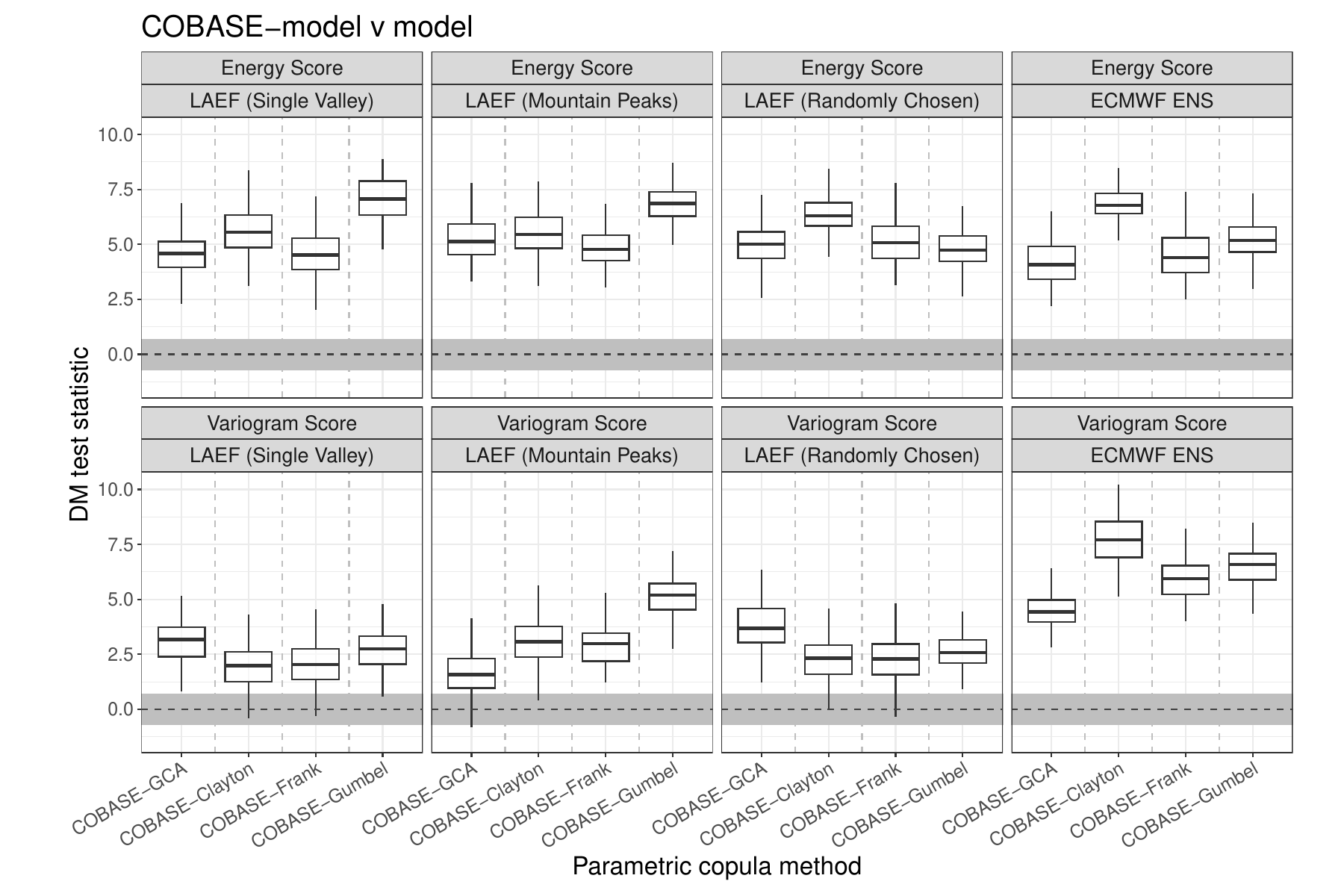}
    \caption{Boxplots of Diebold-Mariano test statistics for the case studies. Here we compare ES and VS of four copula models (baseline) versus their COBASE variants. Values above the gray band indicate better performances compared to the baseline.}
    \label{fig:improvement_plots}
\end{figure}

We also evaluate COBASE with alternative copula families, namely, Clayton, Frank, and Gumbel, to assess its use and evaluate performance under different (non-Gaussian) dependence structures. 
The corresponding DM test results are displayed in Figure~\ref{fig:improvement_plots}. 
These results indicate that the COBASE variants consistently outperform their corresponding standard copula-based implementation. This conclusion is further supported by the ES and VS scores, also reported in Table~\ref{tab:ES_VS}.

\section{Discussion and conclusion}
\label{sec:conclusions}
This paper introduces COBASE, a new multivariate \ep{postprocessing} method that combines the structure of parametric copula models with the flexibility of \ep{nonparametric} methods. 
By decoupling the sampling of margins from the copula-based dependence structure, COBASE addresses a key issue in the copula-based approaches for \ep{postprocessing}, namely, the degradation in forecast quality caused by random sampling from marginal distributions.

A key strength of COBASE is its flexibility. The method is compatible with any parametric copula family, allowing it to adapt to a wide range of dependence structures. 
In addition, it does not impose constraints on how univariate margins are sampled. 
While uniform quantiles are used here, the method is fully compatible with alternative sampling strategies, such as stratified sampling, making it adaptable to different forecasting needs.

Our results from two real-world case studies in Austria and the Netherlands show that COBASE consistently improves upon standard copula-based methods, including GCA, and outperforms Schaake Shuffle. Moreover, COBASE achieves performance comparable to widely used \ep{nonparametric} techniques such as Sim Schaake and ECC, while avoiding their structural limitations and extensive data requirements.

A closer look at the scores in Table~\ref{tab:ES_VS} reveals that COBASE-GCA performs similarly to Sim Schaake, even in the station groups where the latter shows a slight advantage. It is important to note, however, that the comparison is not entirely fair: Sim Schaake and Schaake Shuffle are trained on four years of data, whereas COBASE, like EMOS, uses only a 30-day rolling training window. \ep{This shorter training horizon makes COBASE substantially more flexible and easier to apply in operational contexts. For example, COBASE would be applicable to C-LAEF, whereas Sim Schaake would not be suitable in that case due to the lack of past raw forecasts. While \cite{Lakatos2023} report that increasing the number of historical pool dates for the Schaake Shuffle does not necessarily lead to improved forecast performance, the Schaake Shuffle still relies on historical dependence templates and exhibits weaker average performance in our experiments. In contrast, COBASE achieves improved skill on average while requiring only a short rolling training window, making it a more practical choice in operational settings.} Additionally, COBASE shows robust performance even in relatively high-dimensional settings, such as the ECMWF ENS group with 12 variables, and no dimension-specific tuning.
\ep{Another important aspect is that COBASE, as any parametric copula-based method, can be applied to both exchangeable and nonexchangeable ensemble systems, whereas this may not necessarily hold for ECC and Sim Schaake, depending on the underlying assumptions and, in the latter case, on the chosen similarity criterion. While ECC formally relies on ensemble exchangeability, previous work in \cite{Perrone2020} has shown that nonexchangeability does not necessarily prevent its application, although it requires careful diagnostic analysis to identify potential anomalous dependence structures in the raw forecasts. Such considerations depend on the specific characteristics of the NWP system and are therefore model-dependent. For this reason, we include ECC in this study for comparison purposes. }

The added value of COBASE becomes particularly evident in comparison to that of its corresponding standard copula-based variant. In this case, the performance gap is much more pronounced, confirming that the observed improvements stem from the structure of the COBASE method itself rather than from the specific copula family employed. This supports the main goal of the study: Not to identify the single best-performing copula, but to demonstrate that, across different configurations, COBASE consistently enhances the effectiveness of copula-based \ep{postprocessing}. In this sense, COBASE represents a more efficient, flexible, and broadly applicable alternative to conventional parametric multivariate approaches.

In the future, we plan to further investigate several aspects that could enhance the performance and versatility of the method.
First, the quality of the estimated dependence structure depends strongly on both the choice of copula family and the accuracy of parameter estimation, which can be affected by limited sample sizes. In this study, we focused on temperature, a variable that is approximately Gaussian, so the Gaussian copula (GCA) already provided a good representation of dependence. For other variables, such as wind speed or precipitation, where deviations from Gaussianity are more pronounced or extremes are of greater interest, exploring alternative copula families or model selection strategies may yield further improvements.

Second, future work could aim to make COBASE more adaptive by allowing the dependence structure, and thus the copula family, to vary dynamically with the forecast date or prevailing meteorological regime. Such adaptivity could enable the method to better capture time-varying or state-dependent relationships between variables. However, implementing dynamically varying copulas would also increase computational demands, so developing efficient strategies for adaptive copula selection or updating will be essential for practical applications.

Then, while COBASE avoids the limitations of random marginal sampling, it still relies on random draws from the fitted copula to represent the multivariate structure. Future research could explore alternative multivariate sampling strategies to reduce sampling variability and better capture the characteristics of the dependence structure, especially in the tails of the multivariate distributions. These avenues offer promising directions for further strengthening COBASE's performance across a wider range of postprocessing tasks.

\ep{Finally, an important direction for future work is the development of controlled simulation studies to further assess the performance of COBASE relative to benchmark methods, and to systematically investigate the role of ensemble size and ensemble exchangeability.}

To conclude, taken together, our findings demonstrate that COBASE already provides a simple, flexible, and effective framework for multivariate ensemble \ep{postprocessing}. By combining the strengths of parametric and \ep{nonparametric} approaches, it improves forecast skill without requiring long historical records or raw ensemble data. The COBASE methodology is also particularly well suited for research on \ep{postprocessing} of multivariate dependence, as it allows a clear separation between the evaluation of the dependence structure and the marginals, thereby facilitating a better identification of potential improvements in the dependence component itself. These properties make COBASE particularly well-suited for operational settings with limited data or for newly developed forecasting systems, such as the C-LAEF system in Austria.

Building on this foundation, future work will focus on adjusting the method to higher-dimensional settings, improving copula selection strategies, and enhancing the multivariate sampling step. With these developments, COBASE has the potential to become a robust and scalable solution for a wide range of ensemble \ep{postprocessing} applications.

\ep{\section*{Acknowledgment}
We thank the two anonymous reviewers and the associate editor for their thoughtful comments, which helped improve an earlier version of the manuscript.
}

\ep{\section*{Author Contribution Indication}
Maurits Flos: Conceptualization, Data curation, Formal analysis, Investigation, Software, Validation \\
Bastien Fran\c cois: Conceptualization, Data curation, Investigation, Validation, Visualization, Writing – original draft, Writing – review \& editing \\
Irene Schicker: Data curation, Writing – original draft, Writing – review \& editing \\
Kirien Whan: Conceptualization, Data curation, Investigation, Software, Validation, Visualization, Writing – original draft, Writing – review \& editing \\
Elisa Perrone: Conceptualization, Formal analysis, Investigation, Methodology, Software, Supervision, Validation, Visualization, Writing – original draft, Writing – review \& editing 
}

\section*{Data availability statement}
The data that support the findings of this study are available from the corresponding author upon reasonable request. To facilitate reproducibility, all analysis scripts and mock data that closely replicate the original datasets, allowing reproduction of the figures and tables, are available at \url{https://github.com/elisaperrone/COBASE_github}.

\section*{Conflict of interest statement}
The authors have no conflicts of interest to declare.

\bibliography{references}

\clearpage
\begin{appendix}

\section{Abbreviations}

\begin{table}[htbp]
    \centering
    \setlength{\arraycolsep}{2pt} 
    \renewcommand{\arraystretch}{1.5} 
    \begin{tabular}{c|c}

    \textbf{Method} & \textbf{Label} \\
    \hline
    \hline
    \multicolumn{2}{c}{Univariate postprocessing} \\
    \hline
    \hline
    EMOS (random sampling) & \textit{EMOS-R} \\
    EMOS (uniform quantile sampling) & \textit{EMOS-Q} \\
    \hline
    \hline
    \multicolumn{2}{c}{Multivariate postprocessing} \\
    \hline
    \hline
    Shuffling: & \\
    Sim Schaake + EMOS-R & \textit{SimSSh-R} \\
    Sim Schaake + EMOS-Q & \textit{SimSSh} \\
    Schaake Shuffle + EMOS-Q & \textit{SSh} \\
    ECC + EMOS-Q & \textit{ECC} \\
    \hline
    \hline
    Copula approaches (parametric): & \\ 
    Gaussian Copula &  \textit{GCA} \\
     Clayton Copula &  \textit{Clayton} \\
     Frank Copula &  \textit{Frank} \\ 
     Gumbel Copula &   \textit{Gumbel} \\
    \hline
    \hline
    Copula + Shuffling: & \\
     COBASE - Gaussian Copula + EMOS-Q &  \textit{COBASE-GCA} \\
     COBASE - Clayton Copula + EMOS-Q &  \textit{COBASE-Clayton}\\ 
    COBASE - Frank Copula + EMOS-Q &  \textit{COBASE-Frank} \\
     COBASE - Gumbel Copula + EMOS-Q &  \textit{COBASE-Gumbel}\\ 
    \hline
    \hline
    \end{tabular}
    \caption{Overview of the statistical \ep{postprocessing} methods evaluated in the study, grouped by approach type, with their corresponding labels.}
    \label{tab:method_labels}
\end{table}

\newpage
\section{Scores}
\subsection{CRPS}
\begin{table}[ht]
    \small
    \centering
    \begin{tabular}{lcccccc}
      \hline
    Station & EMOS-R & EMOS-Q & GCA & Gumbel & Clayton & Frank \\ 
      \hline
    Wien Hohe Warte (11035) & 0.8376 & \cellcolor{black!10}\textbf{0.8055} & 0.8417 & 0.8457 & 0.8409 & 0.8548 \\ 
      Patscherkofel (11126) & 1.3846 & \cellcolor{black!10}\textbf{1.3429} & 1.3920 & 1.3903 & 1.3939 & 1.3883 \\ 
      Graz (11290) & 0.9166 & \cellcolor{black!10}\textbf{0.8848} & 0.9210 & 0.9244 & 0.9195 & 0.9260 \\ 
      Pitzal (11316) & 1.0912 & \cellcolor{black!10}\textbf{1.0460} & 1.0831 & 1.0791 & 1.0742 & 1.0838 \\ 
      Innsbruck (11320) & 1.2000 & \cellcolor{black!10}\textbf{1.1515} & 1.1922 & 1.1871 & 1.1946 & 1.2020 \\ 
      Sonnblick (11343) & 0.9650 & \cellcolor{black!10}\textbf{0.9257} & 0.9624 & 0.9640 & 0.9530 & 0.9522 \\ 
      Kolm Saigurn (11344) & 1.3982 & \cellcolor{black!10}\textbf{1.3445} & 1.4146 & 1.3989 & 1.3920 & 1.3938 \\ 
      Rauris (11346) & 1.5313 & \cellcolor{black!10}\textbf{1.4726} & 1.5289 & 1.5421 & 1.5542 & 1.5514 \\ 
      De Kooy (235) & 0.6071 & \cellcolor{black!10}\textbf{0.6030} & 0.6113 & 0.6106 & 0.6133 & 0.6096 \\ 
      Schiphol (240) & 0.5901 & \cellcolor{black!10}\textbf{0.5873} & 0.5957 & 0.5956 & 0.5983 & 0.5973 \\ 
      De Bilt (260) & 0.5772 & \cellcolor{black!10}\textbf{0.5695} & 0.5777 & 0.5786 & 0.5748 & 0.5758 \\ 
      Groningen (280) & 0.6068 & \cellcolor{black!10}\textbf{0.5939} & 0.6026 & 0.6045 & 0.6029 & 0.5987 \\ 
      Vlissingen (310) & 0.6521 & \cellcolor{black!10}\textbf{0.6421} & 0.6544 & 0.6500 & 0.6512 & 0.6522 \\ 
      Maastricht (380) & 0.6325 & \cellcolor{black!10}\textbf{0.6232} & 0.6316 & 0.6310 & 0.6293 & 0.6335 \\ 
       \hline
    \end{tabular}
    \caption{CRPS values for all individual stations and postprocessing methods for 2-meter temperature. To
    avoid redundancy, results are presented for EMOS-R, EMOS-Q, GCA, Gumbel, Clayton and Frank methods only. All the other approaches considered in this study rely on reshuffling the marginals of these methods and produce identical marginal performance. The gray cells indicate the best scores.}
    \label{tab:crps}
    \end{table}

\subsection{ES and VS}
\begin{table}[h]
    \small
    \centering
    \begin{tabular}{|l|c|c|c|c|c|c|c|c|}
    \hline
    \multirow{3}{*}{Method} & \multicolumn{8}{|c|}{Regions and scores} \\
    \cline{2-9}
    \multirow{1}{*}{} &
    \multicolumn{2}{c|}{LAEF (Single Valley)} &
    \multicolumn{2}{c|}{LAEF (Mountain Peaks)} &
    \multicolumn{2}{c|}{LAEF (Randomly Chosen)} &
    \multicolumn{2}{c|}{ECMWF ENS} \\
    \cline{2-9}
     &  ES & VS  & ES & VS  & ES & VS  & ES & VS \\
    \hline
    Raw Ensemble & 7.6462 & 239.5952 & 4.3578 & 57.2592 & 7.3904 & 202.6426 & 3.1251 & 354.8837 \\ 
    \hline
      SSh & 2.6007 & 50.9879 & 2.2322 & 20.0152 & 1.9618 & 19.7619 & 2.7629 & 303.3426 \\ 
      SimSSh-R & 2.6187 & 50.1894 & 2.2837 & 20.1669 & 1.9824 & 19.6075 & 2.7084 & 305.0797 \\ 
      SimSSh & \cellcolor{black!15}\textbf{2.5292} & \cellcolor{black!15}\textbf{47.8356} & 2.2079 & 19.3942 & \cellcolor{black!15}\textbf{1.9325} & \cellcolor{black!15}\textbf{18.6810} & 2.6923 & 299.1751 \\ 
      ECC & 2.6575 & 53.2521 & 2.2130 & 19.5822 & 1.9513 & 19.3800 & 2.6989 & \cellcolor{black!15}\textbf{298.7708} \\ 
      \hline
      GCA & 2.6180 & 50.1523 & 2.2569 & 19.5415 & 1.9860 & 19.8129 & 2.7078 & 303.4243 \\ 
      Clayton & 2.6265 & 50.4588 & 2.2530 & 19.6612 & 1.9862 & 19.4570 & 2.7137 & 303.2181 \\ 
      Frank & 2.6254 & 50.4119 & 2.2593 & 19.7900 & 2.0008 & 19.7574 & 2.7093 & 304.4875 \\ 
      Gumbel & 2.6363 & 50.6686 & 2.2587 & 20.0026 & 1.9843 & 19.5161 & 2.7089 & 303.1059 \\ 
      \hline
      COBASE-GCA & 2.5344 & 48.3626 & \cellcolor{black!15}\textbf{2.2073} & \cellcolor{black!15}\textbf{19.0118} & 1.9363 & 18.8943 & \cellcolor{black!15}\textbf{2.6908} & 299.2300 \\ 
      COBASE-Clayton & 2.5501 & 48.9828 & 2.2080 & 19.1085 & 1.9355 & 18.7195 & 2.6921 & 299.4446 \\ 
      COBASE-Frank & 2.5504 & 48.9279 & 2.2056 & 19.1347 & 1.9344 & 18.7868 & 2.6905 & 299.7904 \\ 
      COBASE-Gumbel & 2.5601 & 49.4700 & 2.2080 & 19.1400 & 1.9345 & 18.7907 & 2.6926 & 299.3518 \\ 
    \hline
    \end{tabular}
      \caption{ES and VS for T2m (LAEF) and T2m and DPT (ECMWF ENS) for the different multivariate postprocessing methods. The gray cells indicate the best scores.}
      \label{tab:ES_VS}
    \end{table}

\end{appendix}

\end{document}